\documentclass[%
 reprint,
 amsmath,amssymb,
 aps,
]{revtex4-2}

\usepackage{graphicx}
\usepackage{dcolumn}
\usepackage{bm}
\usepackage{xcolor}

\usepackage{ulem}

\begin{document}

\preprint{APS/123-QED}

\title{Multiplet Supercurrents in a Josephson Circuit}

\author
{Ethan G. Arnault$^{1,\dagger}$*, John Chiles$^{1,\dagger}$, Trevyn F.Q. Larson$^1$, Chun-Chia Chen$^1$,\\ Lingfei Zhao$^1$, Kenji Watanabe$^2$, Takashi Taniguchi$^2$,\\ Fran\c{c}ois Amet$^3$, Gleb Finkelstein$^1$
\\
\normalsize{$^{1}$Department of Physics, Duke University, Durham, 27701, NC, USA}\\
\normalsize{$^{2}$Advanced Materials Laboratory, NIMS, Tsukuba, 305-0044, Japan}\\
\normalsize{$^3$Department of Physics and Astronomy, Appalachian State University, Boone, 28607, NC, USA}\\
\normalsize{$^\dagger$ These authors contributed equally to this work
\\
$^\ast$To whom correspondence should be addressed; E-mail:  earnault@mit.edu}}

\date{\today}

\begin{abstract}
Multiterminal Josephson junctions are a promising platform to host synthetic topological phases of matter and Floquet states. However, the energy scales governing topological protection in these devices are on the order of the spacing between Andreev bound states. Recent theories suggest that similar phenomena may instead be explored in circuits composed of two-terminal Josephson junctions, allowing for the topological protection to be controlled by the comparatively large Josephson energy. Here, we explore a Josephson circuit, in which three superconducting electrodes are connected through Josephson junctions to a common superconducting island. We demonstrate the dynamic generation of multiplet resonances, which have previously been observed in multiterminal Josephson junctions. The multiplets are found to be robust to elevated temperatures and are confirmed by exhibiting the expected Shapiro step quantization under a microwave drive. We also find an unexpected novel supercurrent, which couples a pair of contacts that are both voltage-biased with respect to the common superconducting island. We show that this supercurrent results from synchronization of the phase dynamics and pose the question whether it should also carry a topological contribution.
\end{abstract}

\maketitle


\section{Introduction}

Josephson junctions under applied voltage bias demonstrate the AC Josephson effect: the phase across the junctions $\varphi$, evolves at an average rate determined by the voltage, $\langle\dot{\varphi}\rangle \propto V$. This behavior could be contrasted with the DC Josephson effect, in which the phase is constant in time and the dissipationless current is carried by the Cooper pairs at zero voltage. The introduction of additional contacts in multiterminal Josephson junctions brings about new possibilities for creating static phase conditions and associated resonances. The most studied case is found at the biasing condition $V_1+V_2=0$. Here, the static phase condition is $\langle \dot{\varphi_1}+\dot{\varphi_2}\rangle=0$, which corresponds to a current of four electrons (one Cooper pair generated in each contact) flowing to the grounded third contact. Without loss of generality, a larger number of Cooper pairs can be mediated in this manner following the condition 
\begin{equation}
nV_1+mV_2=0 
\end{equation}
with integer $n$ and $m$. Resonances such as these have been dubbed ``multiplets'' and have been realized in Josephson bi-junctions~\cite{Cohen2018} and multiterminal Josephson junctions~\cite{Pfeffer2014,KoFan2021,Arnault2021,graziano2022,arnault2022,ohnmacht2023,gupta2024}. While microscopic mechanisms utilizing cross Andreev reflection~\cite{Cohen2018} and hybridized or Floquet-Andreev bound states~\cite{KoFan2021,MelinFloq,MelinBerry,Coraiola2023} can be responsible for these resonances, recent efforts have shown that quartets can also be classically generated through the device's circuit network~\cite{ Melo2021,graziano2022,arnault2022}.

\begin{figure*}[htp]
    \centering
    \includegraphics[width=1.8\columnwidth]{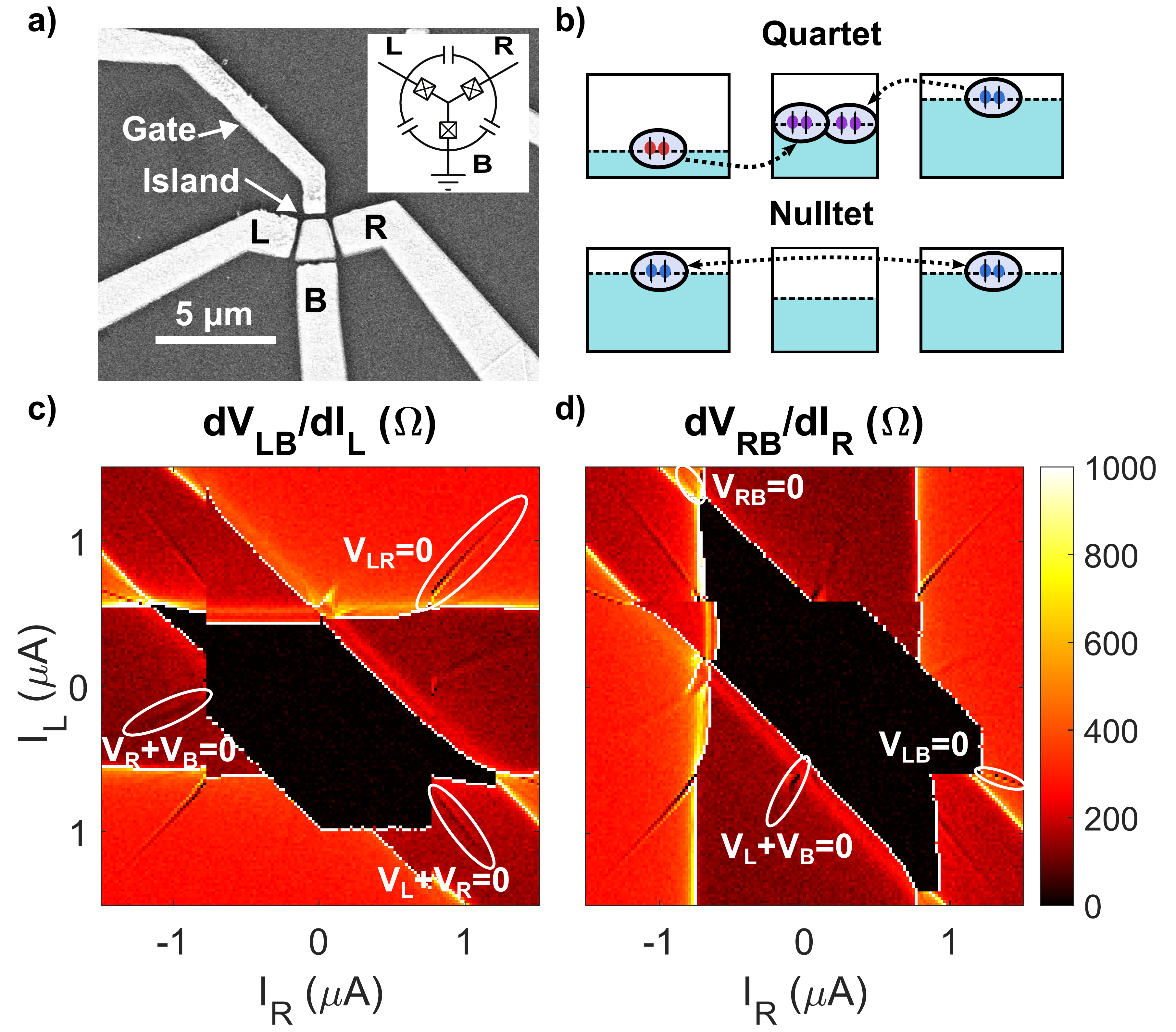}
    \caption {a) SEM image of one of the two island devices studied. Inset: circuit schematic of the device.  b) Schematic of the Cooper pair transport in the device. In the quartet branch two Cooper pairs from distinct contacts form a $4e$, $\sin(\varphi_1+\varphi_2)$ supercurrent. In the nulltet branch a Cooper pair passes through the island to the adjacent contact. c,d) Differential resistance measurements performed between the c) left (L) and bottom (B) as well as the d) right (R) and bottom (B) contacts. Dips in the differential resistance indicate the onset of a superconducting branch. In total, nine fundamental superconducting branches are observed on the device with three different origins.
    }
    \label{fig:Fig1}
\end{figure*}

Recent theory has suggested that multiplet resonances can classically emerge in a device with three contacts connected to a floating superconducting island through a weak link (Fig.~\ref{fig:Fig1}a)~\cite{Melo2021}. Like other classical realizations, this circuit geometry benefits from generating multiplet resonances through the collective behavior of the superconducting circuit. This means the phenomenology is dictated by the Josephson energy, often orders of magnitude larger than the Andreev level spacing, making the multiplets more robust. Further, Ref.~\cite{Melo2021} shows that the quartets have a distinct contribution due to the quantum geometry of the energy landscape. As a result, topological effects may be more more accessible than the features discussed in multiterminal Josephson junctions with a shared normal region \cite{Riwar2016, Eriksson2017, Meyer2017,Xie2017,Xie2018,Gavensky2018,GeomtricTensor, Strambini2016,Vischi2017} Therefore, exploration of these circuit-mediated resonances is a crucial first step to realizing topological effects in Josephson circuits.

In this work, we experimentally realize the island superconducting circuit and show that multiplet resonances do emerge as suggested by theory~\cite{Melo2021}.
The quartets are found to be more robust than previous realizations in multiterminal Josephson junctions~\cite{arnault2022}. We further observe an unexpected novel supercurrent mechanism that couples superconducting contacts across the central island, to which they both are voltage biased. This supercurrent branch, which we dubbed ``nulltet'', demonstrates Shapiro steps at quantized voltage values of $V_1-V_2=\frac{khf}{2e}$, resulting in half-quantized steps for either $V_{1,2}$. Our results indicate that large dynamically stabilized supercurrents can be generated in scalable Josephson circuits. Furthermore, exploration of this prototypical device provides a valuable step towards the realization of topological states in Weyl-Josepshon circuits~\cite{WeylCircuit}. 

\section{Results}

The devices studied here feature either a $1 \times 1$ $\mu$m square superconducting island, or a slightly larger trapezoidal island, connected to three contacts left (L), right (R) and bottom (B) via three separate 500 nm long graphene channels, (Fig.~\ref{fig:Fig1}a). The device's contacts and island are made of sputtered molybdenum-rhenium (MoRe), a superconductor known to form high transparency Ohmic contacts to graphene~\cite{Calado2015,Borzenets2016}. The graphene is etched such that there is no direct coupling between contacts L, R, and B, and all current must flow through the island. Additionally, a fourth  electrode is placed near the island to serve as a gate, but it is left unused in this work.

The device is cooled in a Leiden Cryogenics dilution refrigerator to a base temperature of 60 mK. We apply a backgate voltage of 25 V such that the device is tuned far from the Dirac peak, increasing the critical currents. In the typical measurement, DC biases $I_L$ and $I_R$ are applied to the L and R contacts with respect to the cold-grounded B contact. In Fig \ref{fig:Fig1}, we present the low-frequency differential resistances $dV_{LB}/dI_{L}$ and $dV_{RB}/dI_{R}$ of the trapezoidal island sample.

\begin{figure*}[htp]
    \centering
    \includegraphics[width=2.\columnwidth]{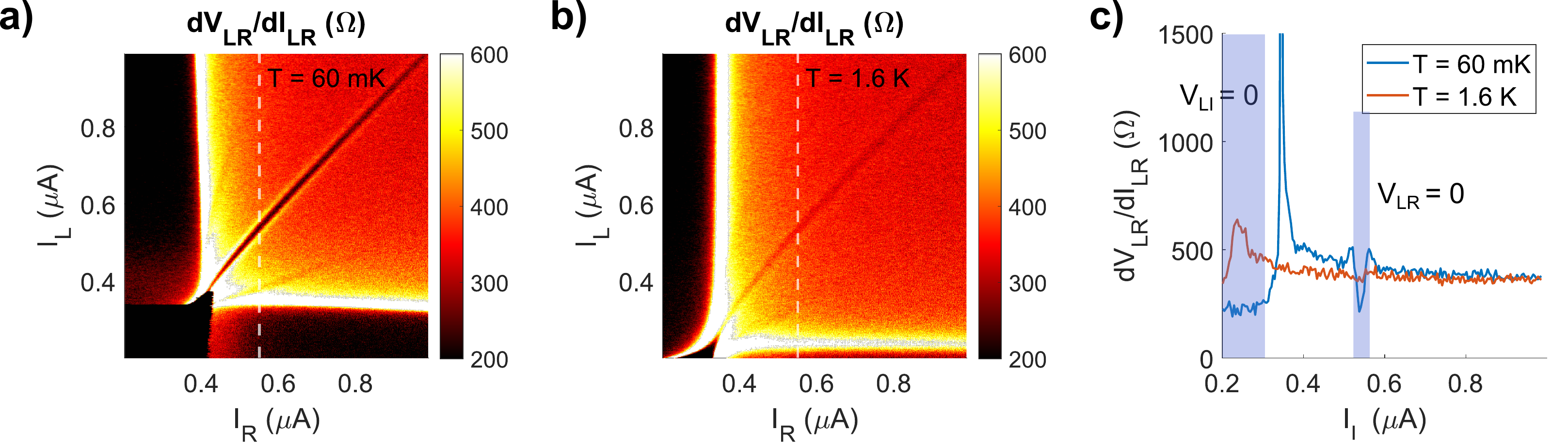}
    \caption {a,b) Maps of differential resistance $dV_{LR}/dI_{LR}$ measured at 60 mK and 1.6 K. In both maps, the nulltet where $V_{L} = V_{R}$ is clearly visible. Additionally, two fainter features appear symmetrically about the nulltet, which we attribute to harmonics of the nulltet. Vertical dashed lines at $I_{R} = 0.55 \mu A$ indicate the location of the cuts plotted in (c). c) Cuts taken from the maps in (a,b). Blue shading highlights two important features: the $V_{LI} = 0$ regime and the $V_{L} = V_{R}$ nulltet.
}
    \label{fig:Fig2}
\end{figure*}

\begin{figure*}[htp]
    \centering
    \includegraphics[width=2\columnwidth]{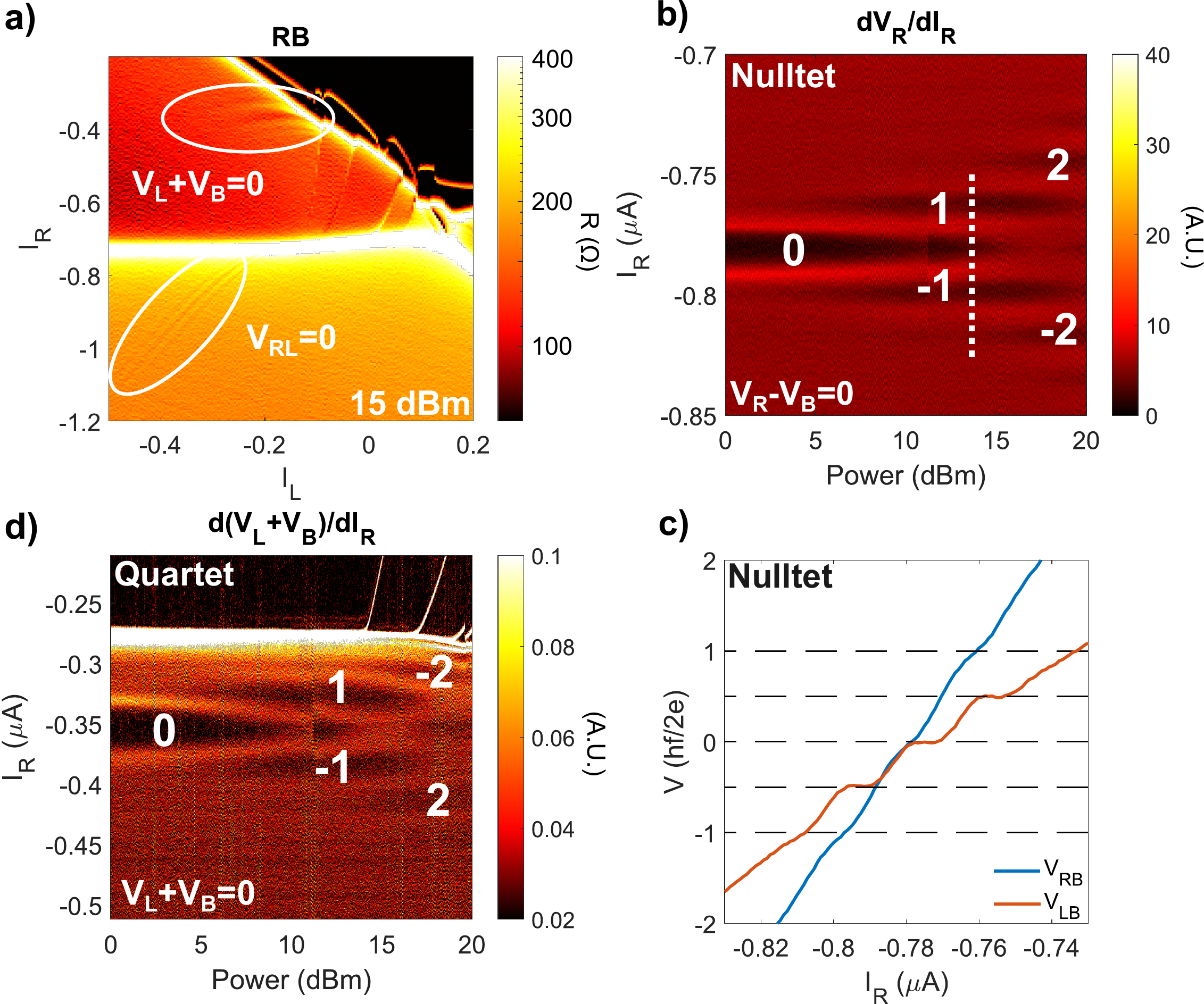}
    \caption {a) A zoom-in map of the numerical differential resistance $dV_{RB}/dI_{R}$ with an applied power of 15 dBm at 5.2GHz. Both the quartet branch $V_L+V_B=0$ and the non-local superconducting branch $V_{RL}=0$ produce parallel lines of constant voltage known as Shapiro steps. b) Numerical differential resistance $dV_R/dI_R$ for $I_L=$ 1.5 $\mu$A against applied RF power and DC bias along the nonlocal supercurrent line $V_{RB}=0$. The supercurrent exhibits integer Shapiro steps following $V_R=\frac{nhf}{2e}$ for $n$ as an integer. c) $V_{RB}$ and $V_{LB}$ taken at 14 dBm showing clear integer and fractional Shapiro steps respectively. In $V_{LB}$, the voltage is effectively being probed between the island and the ground. Therefore, the Shapiro steps reflect the behavior of the junction formed between the island and the bottom contact. The voltage values take on clear half integer steps, indicating that each junction forming the nonlocal supercurrent consists of a half step. d) Numerical differential resistance $d(V_L+V_B)/dI_R$ power bias map of the $V_L+V_B=0$ quartet line taken at $I_L$ = -0.21 $\mu$A. Steps emerge for $V_L+V_B = \frac{nhf}{2e}$ for $n$ an integer - consistent with theory (see supplementary information).
    }
    \label{fig:Shapiro}
\end{figure*}

Three large superconducting branches are observed 
in the two maps, each corresponding to a supercurrent coupling the island to one of the three contacts. These branches appear as vertical (L), horizontal (R), and 45$^{\circ}$ (B) bands, corresponding to conditions $|I_L|<I^{(C)}_L$, $|I_R|<I^{(C)}_R$, and $|I_L+I_R|<I^{(C)}_B$. (Here, $I^{(C)}_{L,R,B}$ indicate the critical current of the corresponding contact to the island.)   
The quartet lines in this sample geometry appear as narrow resonances that fall within these main branches. For example, the $V_L+V_R=0$ quartet condition results in $I_L+I_R \approx 0$, and therefore it falls within the B superconducting branch.

Between each of the large branches (L, R, B), additional small superconducting resonances appear that correspond to a supercurrent induced between the contacts (for example $V_L-V_R=0$), while both contacts develop a finite voltage to the island and the third contact. 
The width of the superconducting island in our sample (microns)  greatly exceeds the coherence length of MoRe alloy (a few nm), and therefore these supercurrents cannot be attributed to elastic cotunneling of quasiparticles ~\cite{Matsuo2022}. Instead, we argue that these superconducting branches are generated through the device's circuit. We refer to these resonances as ``nulltets'', because, in contrast to quartets, no net Cooper pairs are exchanged between the contacts and the island. In total, six main narrow resonances are observed in the device, corresponding to conditions $V_i=\pm V_j$, where $i,j$ correspond to L, R, and B. For clarity, in Fig.~\ref{fig:Fig1}c, we provide a map of the resonances observed between the L and B contacts.

In addition to these 9 superconducting branches, we also observe fainter supercurrents (see Fig.~\ref{fig:Fig2} for a higher contrast). It is important to emphasize that these supercurrents are not the ``sextets" previously observed in Josephson bi-junctions~\cite{Cohen2018}, as they appear with opposite sign in their voltage relation. Indeed, these superconducting branches are harmonics of the nulltet, further confirming that the nulltet supercurrents do not originate from a trivial connection between the two leads.

The multiplets correspond to harmonics generated by the nonlinearity of the circuit, and therefore should be robust to elevated temperature. In Fig.~\ref{fig:Fig2}, we show the robustness of the resonances in temperature. We find that the suppressed region of resistance persists up to at least 1.6 K for the nulltet branch, and even some traces of the nulltet's higher harmonic are still apparent in Fig.~\ref{fig:Fig2}b. In the supplementary information, we show that the resonances in the square island device persist even up to 2.2 K. As a comparison, a critical current of $\sim 50$ nA (comparable to the width of the resonances)  corresponds to the Josephson energy of 1 K and would be washed away at even lower temperatures. Instead, here the temperature scale controlling the suppression of the resonances is likely determined by the Josephson energies of the individual junctions, which are several Kelvin. In Ref.~\cite{arnault2022}, we have also observed robust quartet supercurrents, which disappeared only by 1.8 K. Notably, the quartet supercurrents are now observed at resistances higher than in the previous works~\cite{KoFan2021,graziano2022,arnault2022}, including the prediction of our model~\cite{arnault2022} describing a $\Delta$-shaped three terminal circuit
with comparable capacitances. We suspect that this speaks to the more robust nature of the circuit design presented in this paper.
Specifically, instead of mutual Joule heating through a shared normal region, each junction is thermally isolated by the central superconducting island. 

\begin{figure*}[htp]
    \centering
    \includegraphics[width=2\columnwidth]{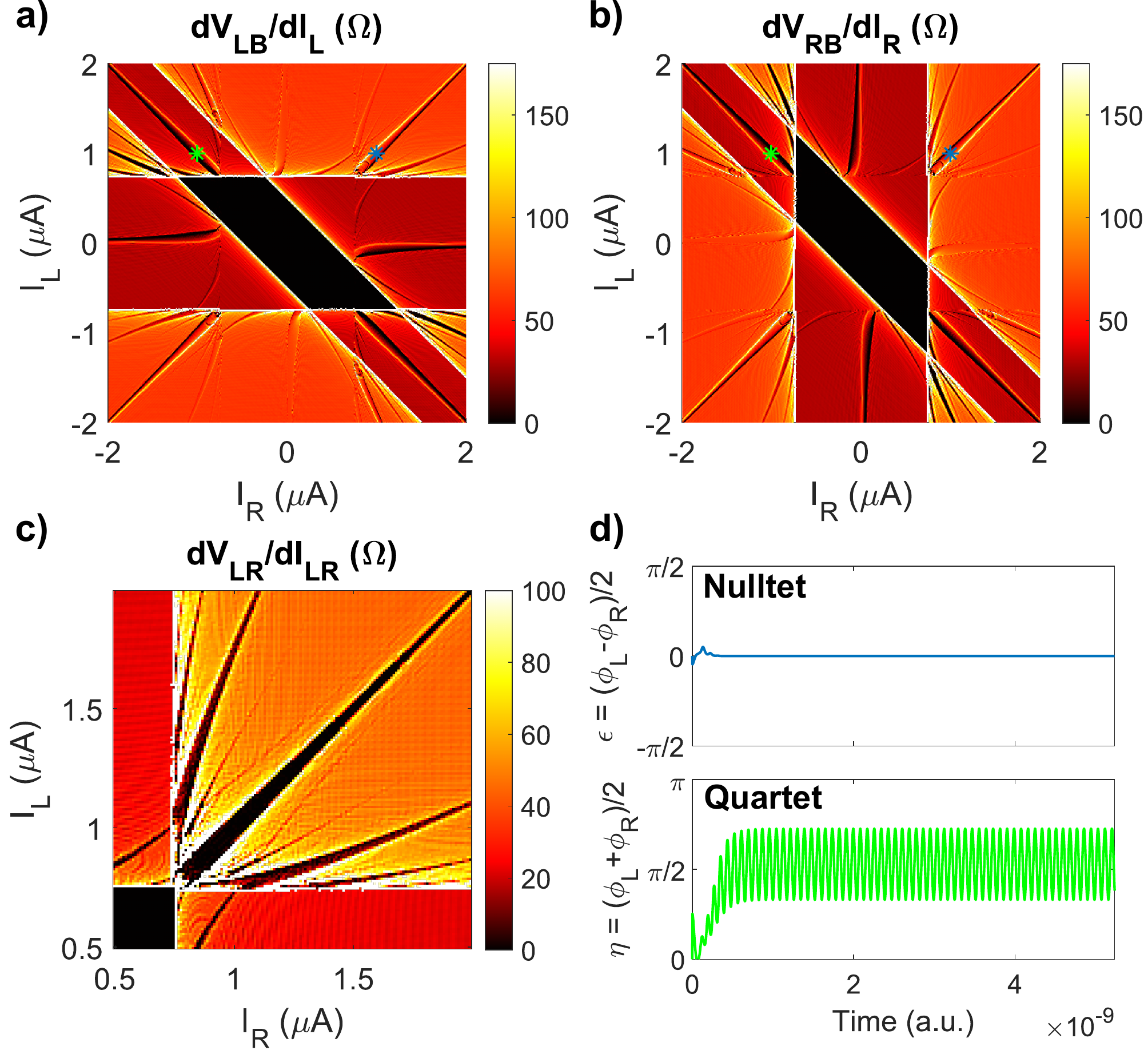}
    \caption {a-c) Numerically calculated differential resistances as a function of bias of the a) left bottom, b) right bottom and c) left right junctions. These maps reproduce all supercurrents observed in the experiment. d) Phase trajectories taken in the (top) nulltet and (bottom) quartet superconducting branches. Color coded stars in panels (a) and (b) denote the location in bias. Both trajectories stabilize around an integer multiple of $\pi/2$ as predicted by the Kapitza model (see supplementary). The quartet feature has weak oscillations due to a small deviation in the Kapitza model due to the island phase.
    }
    \label{fig:Sim}
\end{figure*}

The lack of sensitivity to elevated temperature allows us to comfortably apply a microwave drive (which can elevate the junction to temperatures on the order of a Kelvin~\cite{quasiparticleHeating}) and study the Shapiro steps of each branch. Shapiro steps are the result of mode locking between the superconducting phase and an external drive. This locking produces a quantized voltage proportional to the drive frequency, $f$. The map of the Shapiro steps depends on the CPR of the junction~\cite{Rokhinson2012,Snyder2018} as well as the broader circuitry of the device~\cite{Sullivan1970,Russer1972,Trevyn}.

In order to generate the Shapiro steps, we drive the sample by a 5.2 GHz microwave tone. The microwave excitation applied at room temperature by the signal generator passes through several attenuators, reducing the power reaching the sample by at least 5 orders of magnitude. Since the delivered power varies as a function of frequency, we quote only the power applied by the signal generator. Fig.~\ref{fig:Shapiro}a shows the averaged numerical derivative of the DC voltages measured between the R and B contacts of the square island sample under applied microwave excitation. As power is increased, additional features emerge which run parallel to all the superconducting branches. These additional branches can be ascribed to the generation of Shapiro steps in a multiterminal sample~\cite{Arnault2021}. 

We examine the Shapiro steps of the nulltet branch corresponding to $V_R = V_B$. $I_L$ is fixed at 1.5 $\mu$A, as we sweep $I_R$ and applied power, while measuring $V_{RB}$. Note that in this configuration, the L and B contacts effectively trade places as compared to the L-R nulltet presented in Fig.~\ref{fig:Fig1},~\ref{fig:Fig2}. The nulltet Shapiro steps exhibit Bessel-like behavior (Fig.~\ref{fig:Shapiro}b). Unlike the Shapiro steps of the individual junctions~\cite{Arnault2021,Trevyn}, these steps do not overlap and are therefore free from hysteresis. In Fig.~\ref{fig:Shapiro}c, we plot $V_{RB}$ measured vs $I_R$ at $P_{RF}=$ 14 dBm (white line in Fig.~\ref{fig:Shapiro}b) and find that the voltage steps are integer. From here, the $V_{LB}$ junction is measured in the same fashion. When $I_L$ is fixed, the left junction's voltage is fixed with respect to the island, therefore this measurement effectively probes the voltage between the island and the bottom contact (up to a constant offset from the left junction). At the same value of the applied RF power, we find that the steps of $V_{LB}$ are spaced by $\frac{1}{2} \frac{hf}{2e}$. This indicates that the $V_{RB}$ voltage steps are evenly distributed between the R and B contacts and the island. This distribution of voltages is not caused by both junctions natively exhibiting half steps -- fractional steps due to CPR in a single junction are suppressed in single graphene junctions of comparable dimensions~\cite{Trevyn}. We therefore conclude that the half-integer step voltages emerge from the collective dynamics of both junctions.

Next, we focus on the quartet branch which corresponds to the voltage condition $V_L+V_B=0$ at zero power. In Fig.~\ref{fig:Shapiro}d, we plot the numerical derivative of $V_L+V_B$ vs $I_R$ and RF power, while keeping $I_L$ set at $-0.21 \mu$A. In the supplementary we show cross-sections at various powers. The voltage steps follow $V_L+V_B=\frac{nhf}{2e}$ with integer $n$. These quantized voltage values adhere to the analytical solution, which follows a modified version of the conventional derivation for a single Josephson junction~\cite{Tinkham} (see Supplementary information). We find that these quantized values are expected for the quartet branch with a CPR of $ I^0(\varphi_1,\varphi_2)=I^0_{1,1}\sin(\varphi_1+\varphi_2)$.

Finally, we perform numerical calculations showing that our results can be reproduced using the resistively and capcitively-shunted Junction (Stewart McCumber) model. Fig.~\ref{fig:Sim}a,b and Fig.~\ref{fig:Sim}c (zoomed in) reproduce maps similar to those in Fig.~\ref{fig:Fig1} and Fig.~\ref{fig:Fig2}. We are able to simulate trivial, quartet, nulltet and higher harmonics of the dynamical supercurrents. Notably, in order to accurately reproduce the features, we must include a large capacitance between each of the contacts. Physically, it describes the large bonding pads which serve as a capacitive shunt on the order of several pF~\cite{Trevyn}. Additionally, we find that we can safely neglect the capacitive coupling of the island to the leads. Conceptually, the negligible capacitance of the island allows its phase to rapidly evolve, which may enable the observed dynamical effects.

To better understand the origin of these dynamical supercurrents, in Fig.~\ref{fig:Sim}d we plot the phase evolution of the superconducting contacts in both the nulltet (top panel) and quartet (lower panel) branches. We have previously discussed the origins of the quartet resonances in Josephson junctions connected in a $\Delta$ circuit~\cite{arnault2022}. The state of that system was characterized by two phase differences, and the quartets corresponded to the ``phase particle'' gliding through a higher dimensional energy landscape  along the diagonals determined by the conditions $\varphi_L + \varphi_R = \mathrm{const}$. 

The present ``star circuit'' is characterized by three phase differences (one per junction). As we show in the supplementary, we can consider combinations of phases, $\eta = \phi_L + \phi_R$ and $\epsilon = \phi_L-\phi_R$, and find that both the quartet and the nulltet resonances have similar origins. Here, the resonances follow a trajectory in the phase space that is stabilized via a mechanism analogous to the Kapitza pendulum. Notably, the superconducting phase of the quartet branch slightly deviates from the Kapitza model. Here, the superconducting phase of the island perturbs the system and results in small oscillations about $\eta = \pi/2$.

In summary, we fabricated a Josephson circuit described in Ref.~\cite{Melo2021} which consists of a superconducting island contacted by three tunable graphene Josephson junctions. The device exhibits nine fundamental superconducting resonances, with three different origins. 1) A large, trivial resonance associated with a supercurrent formed between the island and each of the contacts. 2) A nonlocal supercurrent mediated between two contacts through the superconducting island, whose origin is similar to the dynamical stabilization of quartet supercurrents in three terminal Josephson junctions~\cite{arnault2022}. 3) A quartet, which is born from the circuit geometry and was predicted in Ref.~\cite{Melo2021}. Remarkably, these multiplet supercurrents are more robust than previous realizations in three terminal Josephson junctions, as they exist in a more resistive device and persist at temperatures exceeding 2 K. This allows us to confirm the resonances here are indeed quartets by examining their Shapiro steps, finding the expected integer steps for $V_1+V_2=\frac{nhf}{2e}$.  Collectively, these results demonstrate that creating circuit-driven macroscopic phase coherent transport is possible in systems beyond multiterminal Josephson junctions. 

Our findings open new pathways towards several emerging technologies. This device geometry has already demonstrated a tunable superconducting diode effect~\cite{Chiles2023}, and engineering multiplet states could further enhance the non-reciprocity ratio beyond simple biasing techniques. Further, multiplet supercurrents in these devices might be useful for creating topologically protected superconducting qubit architectures like $\cos2\varphi$ qubits~\cite{Gladchenko2009,Smith2020,Melo2021, Marcus2020}. Our findings are in principle agnostic to the type of the utilized weak link and therefore graphene can be replaced by tunnel junctions, opening a route to explore these technologies in a scalable platform which will compatible with current superconducting qubit fabrication techniques. Finally, in addition to promising technological directions, our work establishes a new beginning for exploring topological states in superconducting circuits, as predicted in Refs.~\cite{WeylCircuit,Melo2021}.

\section{Acknowledgements}

We thank Andre Melo, Valla Fatemi, and Anton Akhmerov for helpful discussions. Transport measurements by E.G.A., J.C., T.F.Q.L. and C.C., lithographic fabrication and characterization of the samples by E.G.A., L.Z. and F.A., and data analysis by E.G.A., F.A. and G.F., were supported by Division of Materials Sciences and Engineering, Office of Basic Energy Sciences, U.S. Department of Energy, under Award No. DE-SC0002765. K.W. and T.T. acknowledge support from JSPS KAKENHI Grant Number JP15K21722 and the Elemental Strategy Initiative conducted by the MEXT, Japan. T.T. acknowledges support from JSPS Grant-in-Aid for Scientific Research A (No. 26248061) and JSPS Innovative Areas “Nano Informatics” (No. 25106006). This work was performed in part at the Duke University Shared Materials Instrumentation Facility (SMIF), a member of the North Carolina Research Triangle Nanotechnology Network (RTNN), which is supported by the National Science Foundation (Grant ECCS-1542015) as part of the National Nanotechnology Coordinated Infrastructure (NNCI).

\clearpage

\section{Supplementary Information}

\subsection{Derivation of Shapiro Step Voltages Along a Quartet Branch}

In order to find the expected voltage quantization values we follow the standard approach by taking a time varying voltage:
\begin{align*}
    V(t)=V_{DC}+V_{AC}\cos\omega t
\end{align*}
and insert this into the Josephson relation:
\begin{multline*}
    \varphi=\frac{2e}{\hbar}\int V_{DC}+V_{AC}\cos\omega t dt\\=\varphi_0 +\frac{2e}{\hbar} (V_{DC}t+\frac{1}{\omega}V_{AC}\sin\omega t)
\end{multline*}
This phase can be inserted into the current phase relation, which is assumed to be of the form $I_c\sin(\varphi_1+\varphi_2)$ with time varying voltages $V_1(t)$ and $V_2(t)$ evolving $\varphi_1$ and $\varphi_2$ respectively. It is convenient to rewrite this equation as an exponential yielding:
\begin{multline*}
    I=I_c \mathrm{Im}[\exp (i(\varphi_0 +\frac{2e}{\hbar} (V_{DC,1}t+\frac{1}{\omega}V_{AC,1}\sin\omega t\\+V_{DC,2}t+\frac{1}{\omega}V_{AC,2}\sin\omega t)))]
\end{multline*}
Here, we assume that the drive frequency $\omega$ is the same for each contact. Taking $\alpha=\frac{2e(V_{DC,1}+V_{DC,2})}{\hbar}$, $\beta=\frac{2e(V_{AC,1}+V_{AC,2})}{\hbar\omega}$ and $\gamma=\omega t$ the equation reads:
\begin{align*}
    I=I_c \mathrm{Im}[\exp (i(\varphi_0 +\alpha t+\beta\sin\gamma)))]
\end{align*}
\begin{align*}
    =I_c \mathrm{Im}[\exp (i(\varphi_0 +\alpha t)\exp(i\beta\sin\gamma)))]
\end{align*}
It can be noticed that the second exponential is equivalent to the Bessel functions. Simplifying further:
\begin{align*}
    I = I_c \sum^{\infty}_{k=-\infty} (-1)^k J_k(\beta)\sin(\varphi_0+\alpha t-k\gamma)
\end{align*}
Relating this to the RSJ equation (excluding capacitance), one finds the current across the junction is $(V_{DC,1}+V_{DC,2})/R_n$, unless $V_{DC,1}+V_{DC,2}=\frac{k\hbar \omega}{2e}$, where the current will spike with an amplitude dictated by the Bessel function. It is useful to notice that along the quartet line $V_{DC,1}=-V_{DC,2}$, therefore biasing in this case is a measure of how incommensurate the biasing voltages are. We define this value as $V_{DC,1}+V_{DC,2}=V_q$. Therefore, the current spikes when $V_q$ is an integer multiple of $\frac{\hbar \omega}{2e}$. 

\begin{figure}[htp]
    \centering
    \includegraphics[width=\columnwidth]{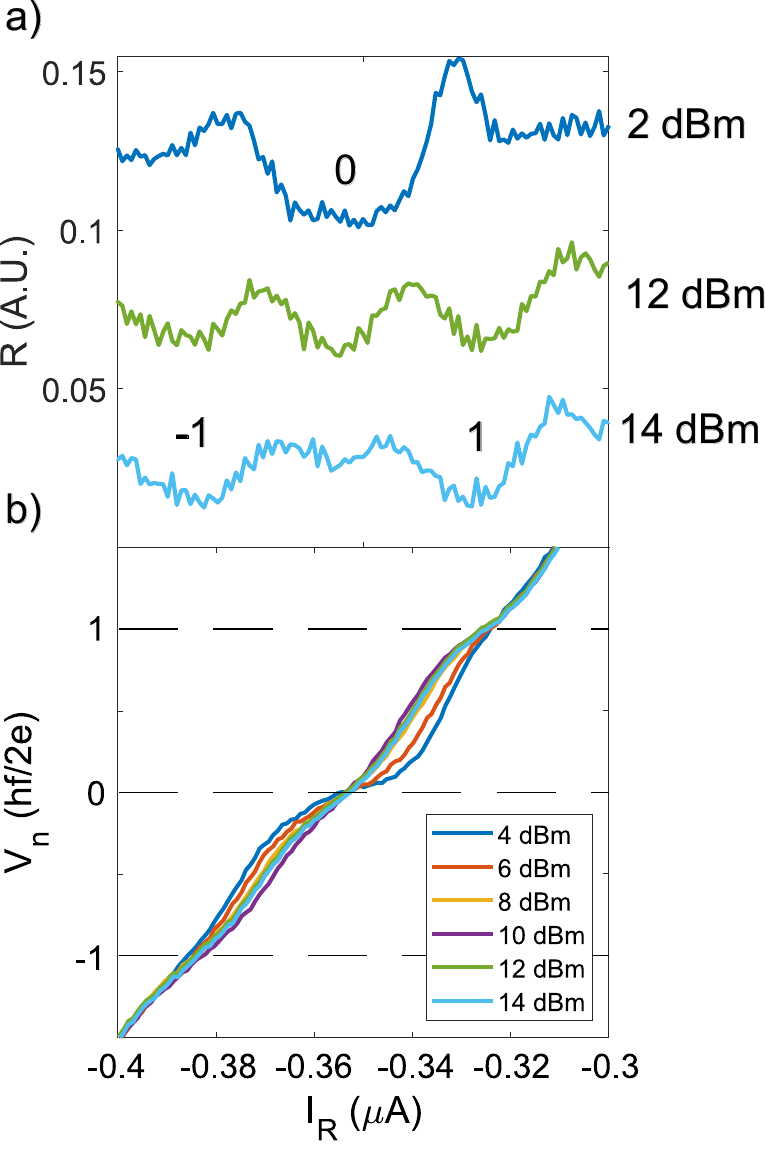}
    \caption {a) Locally averaged cuts from Fig. 3d taken at 2 dBm, 12 dBm, and 14 dBm. Each cut averages the signfrom $\pm$0.4 dBm in power. b) Locally averaged quartet voltage cuts taken from main Fig. 3d at several different microwave powers.
    Shapiro steps follow $V_L + V_B = \frac{nhf}{2e}$ as predicted by theory.}
    \label{fig:QuartetShapiro}
\end{figure}

In agreement with this theory, Fig.~\ref{fig:QuartetShapiro} shows differential and voltage cuts of the Shapiro steps shown in Fig.~\ref{fig:Shapiro}. Shapiro steps take the form $V_L + V_B = V_q = \frac{nhf}{2e}$. 

\subsection{Mathematical Modeling}

To better understand the mathematical basis of this circuit geometry, we construct a simple toy model using the Stewart-McCumber theory. For simplicity, we model a symmetric network where all Josephson junctions have the same critical current, and are each shunted by the same resistance. We neglect the junction capacitance. Additionally, we take into account the large capacitance $C$ between the bonding pads and ground. We define $\omega_{0}^{2}=\frac{2eI_{C}}{\hbar C}$,  left and right input currents $\tilde{i}_{L}$ and $\tilde{i}_{R}$ normalized by the critical current of the junctions, and use the dimensionless time variable $\tau\equiv\omega_{0}t$. From Kirchhoff laws we obtain:

\begin{align*}
\tilde{i}_{L}&=(2\ddot{\varphi}_{L}-\ddot{\varphi}_{R})+\sin(\varphi_{L}-\varphi_{i})+\beta(\dot{\varphi}_{L}-\dot{\varphi}_{i})\\
\tilde{i}_{R}&=(2\ddot{\varphi}_{R}-\ddot{\varphi}_{L})+\sin(\varphi_{R}-\varphi_{i})+\beta(\dot{\varphi}_{R}-\dot{\varphi}_{i})\\
\sin(\varphi_{i})&=\sin(\varphi_{L}-\varphi_{i})+\sin(\varphi_{R}-\varphi_{i})\\
&+\beta(\dot{\varphi}_{L}+\dot{\varphi}_{R}-3\dot{\varphi}_{i})
\end{align*}

The first two equations could be combined:

\begin{flalign*}
\tilde{i}_{L}+\tilde{i}_{R}&=(\ddot{\varphi}_{L}+\ddot{\varphi}_{R})+\sin(\varphi_{L}-\varphi_{i})+\sin(\varphi_{R}-\varphi_{i})
\\&+\beta(\dot{\varphi}_{L}+\dot{\varphi}_{R}-2\dot{\varphi}_{i})\\
\tilde{i}_{L}-\tilde{i}_{R}&=3(\ddot{\varphi}_{L}-\ddot{\varphi}_{R})+\sin(\varphi_{L}-\varphi_{i})-\sin(\varphi_{R}-\varphi_{i})\\
&+\beta(\dot{\varphi}_{L}-\dot{\varphi}_{R})
\end{flalign*}

We now define $\eta\equiv\frac{\varphi_{L}+\varphi_{R}}{2}$ and $\epsilon\equiv\frac{\varphi_{L}-\varphi_{R}}{2}$. Using trigonometric identities, we get:

\begin{align}
\frac{\tilde{i}_{L}-\tilde{i}_{R}}{2}&=3\ddot{\epsilon}+\cos(\eta-\varphi_{i})\sin(\epsilon)+\beta\dot{\epsilon}\\
\frac{\tilde{i}_{L}+\tilde{i}_{R}}{2}&=\ddot{\eta}+\sin(\eta-\varphi_{i})\cos(\epsilon)
+\beta(\dot{\eta}-\dot{\varphi}_{i})
\end{align}

Nulltet/quartet resonances occur when $V_{L} = \pm V_{R}$ and $\langle \dot{\eta} \rangle$ or $\langle \dot{\epsilon} \rangle$ is equal to zero. In junctions with similar normal resistances, this happens for the biasing condition close to $I_{L} = \pm I_{R}$. 

For reasons that will be clear soon, we begin by examining equation (2), which corresponds to the nulltet supercurrent. The nulltet appears for voltages near the condition $V_L = V_R$, which means $\dot{\phi_L} = \dot{\phi_R}$. Therefore, we see that $\dot{\epsilon} \sim 0$ and $\dot{\eta} \sim \dot{\phi_{L,R}}$. If the system is driven by large enough current, $\eta$ will evolve rapidly, and $\eta-\phi_i$ can be effectively replaced by $\omega t$. 

This situation is similar to the case of quartet in a three-terminal Josephson junction geometry, as considered in our work~\cite{arnault2022}. Following the discussion in~\cite{arnault2022}, eq. (5) simplifies to the equation for the Kapitza's inverted pendulum. Here, $\omega t$ rapidly evolves in time, driving $\epsilon$ to be stable about 0 or $\pi$. The absence of any deviation from the Kapitza model enables a complete stabilization of $\epsilon$, as evidenced by the lack of ringing in the top panel of Fig.~\ref{fig:Sim}d. Therefore, the nulltet supercurrents are dynamically identical to the quartet supercurrents in a three-terminal Josephson junction, with the roles of $\eta$ and $\epsilon$ being interchanged~\cite{arnault2022}. 

On the other hand, equation (3), which corresponds to the quartet, simplifies to a similar equation with a notable deviation. The presence of $\varphi_{i}$ in the sine term prevents an ideal simplification for $\eta$. This causes two significant changes compared to the nulltet case: $\eta$ is not fully stabilized and shows non-decaying oscillations; these oscillations are centered around $\pi/2$, not $0$ or $\pi$, as shown in Fig.~\ref{fig:Sim}d.

We now proceed to show that both of these results could be obtained from an analytic model. Similarly to the simpler cases, we assume that bias $\tilde{i}_{L}-\tilde{i}_{R}$ is large enough so that $\epsilon$ can be replaced with $\omega t$~\cite{arnault2022}. An analytic solution can be found for eq. (3) if we additionally assume that $\varphi_{i}$ performs near-sinusoidal oscillations, $\phi_i=C\cos(\omega t - \phi)$. This is a natural assumption because the quartet line falls inside the superconducting branch of the bottom junction, so $\langle \dot{\phi_i} \rangle =0$. This assumption has been further verified in our modeling; we do not assume $C$ to be small. 

We then introduce $\theta=\eta-\phi_i$. 
Using reduced time and frequency units, equation (6) can then be rewritten as:
\begin{align*}
\ddot{\theta}+\beta\dot{\theta}+\cos(\omega t)\sin(\theta)=C\omega^{2}\cos(\omega t-\phi)+D
\end{align*}
where we relabeled $D=\frac{\tilde{i}_{L}+\tilde{i}_{R}}{2}$. This equation is equivalent to Kapitza's inverted pendulum problem with no gravity and with a periodic transverse drive in addition to the vertical oscillations of the pivot. For simplicity, we first discuss the solution in the absence of damping. Let us assume $\theta=\theta_{S}+A\cos(\omega t)+B\sin(\omega t)$, where and A and B are slowly varying compared to $\omega$. This ansatz yields the following equations:

\begin{align*}
\ddot{\theta_{S}}+\frac{A}{2}\cos(\theta_{S})=D\\
\ddot{A}-\omega^{2}A+2\dot{B}\omega+\sin(\theta_{S})=\omega^{2}C\cos(\phi)\\
-2\dot{A}\omega+\ddot{B}-B\omega^{2}=\omega^{2}C\sin(\phi)
\end{align*}

We then only keep the highest order terms in $\omega$ and get:
\begin{align*}
B&=-C\sin(\phi)\\
A&=\left(-C\cos(\phi)+\frac{\sin(\theta_{S})}{\omega^{2}}\right)
\end{align*}

We inject this in the first equation for $\theta_S$:
\begin{align*}
\ddot{\theta}_{S}+\frac{1}{2}\left(\frac{1}{\omega^{2}}\sin(\theta_{S})-C\cos(\phi)\right)\cos(\theta_{S})=D
\end{align*}

We define $\theta_{S}=\frac{\pi}{2}+u$ and the corresponding effective potential is:

\begin{align*}
U_{eff}&=-\frac{1}{8\omega^{2}}\cos(2\theta_{S})-\frac{C}{2}\cos(\phi)\sin(\theta_{S})\\
&=\frac{1}{8\omega^{2}}\cos(2u)-\frac{C}{2}\cos(\phi)\cos(u)
\end{align*}

We see that the equilibrium position around $\theta_{S}=\pi/2$ (i.e. $u=0$) is stable as long as:

\begin{align*}
C>\frac{1}{\omega^{2}\cos(\phi)}
\end{align*}

The equilibrium position will deviate from $\pi/2$ in the presence of a finite bias $D$, but the phase will be stationary as long as:

\begin{align*}
D< \max\left(\frac{C\cos(\phi)}{2}\sin(u)-\frac{1}{4\omega^{2}}\sin(2u)\right)
\end{align*}

At high frequency/large $C$, this simplifies to 

\begin{align*}
D=\frac{\tilde{i}_{L}+\tilde{i}_{R}}{2} < \frac{C\cos(\phi)}{2}
\end{align*}

These considerations indicate that the solution with $\eta$ oscillating around the equilibrium value of $\pi/2$ (Figure~4d) is not an artifact. Instead, it is stable for a range of parameters determined by the amplitude and phase ($C$ and $\phi$) of the second drive, which physically is provided by the phase of the central island, $\phi_i=C\cos(\omega t - \phi)$. Note that in the nulltet case, the effect of damping is to cause oscillations to converge towards the equilibria obtained from Kapitza's canonical derivation. In the quartet case, however, oscillations are driven by a right hand side which explain why they do not decay. In this case, damping will just cause a phase shift and reduction in the amplitude of the oscillations around the equilibrium, but they do not decay.

\subsection{Square Island Thermal Response}

In addition to the maps provided in Fig.~\ref{fig:Fig2}, we provide the the bias-bias maps of the 1 $\mu$m by 1 $\mu$m square superconducting island sample. We observe that the harmonic supercurrents persist at least up to 2.2 K.

\begin{figure}[htp]
    \centering
    \includegraphics[width=0.75\columnwidth]{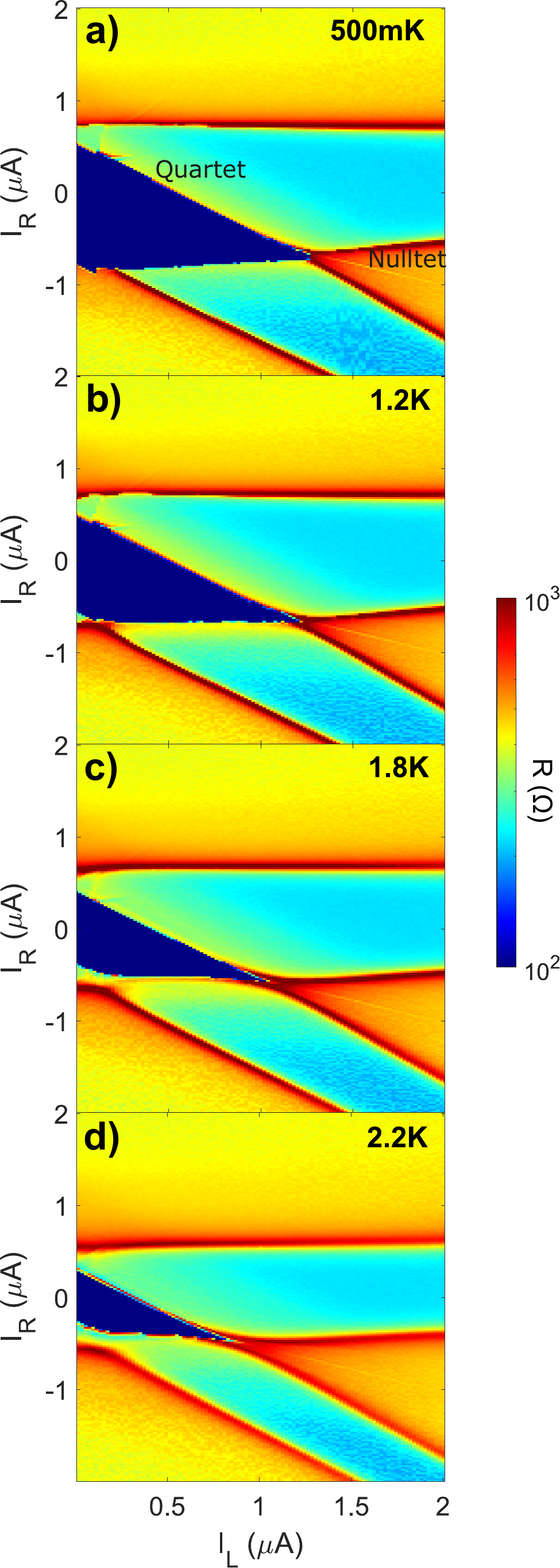}
    \caption {Bias-bias maps of the second, square device as a function of temperature. We find that the quartet and nulltet supercurrents persist up to at least 2.2 K.
    }
    \label{fig:Therm}
\end{figure}

\subsection{Simulation Parameters}

In the main text, we perform RCSJ simulations modeling the effects of the circuit. In Table 1, we provide the values used in Fig.~\ref{fig:Sim}.

\begin{center}
\begin{tabular}{ |c|c|c|c|c|c|c|c|c| } 
 \hline
 $I_{c,L}$ & $I_{c,R}$ & $I_{c,B}$ & $R_{L}$ & $R_{R}$ & $R_{B}$ & $C_{LR}$ & $C_{RB}$ & $C_{LB}$ \\\hline  
 750 nA & 750 nA & 500 nA & 30 $\Omega$ & 30 $\Omega$ & 30 $\Omega$ & 250 fF & 250 fF & 250 fF \\ 
 \hline
\end{tabular}
Table 1. Parameters used in main text Fig.~\ref{fig:Sim}.
\end{center}

\clearpage

\bibliography{apssamp}

\end{document}